\documentclass[twocolumn,aps,amsmath,amsfonts,showpacs]{revtex4}
\usepackage[colorlinks=true]{hyperref}

\newcommand{\be}{\begin{equation}} \newcommand{\ee}{\end{equation}}
\newcommand{\bea}{\begin{eqnarray}} \newcommand{\eea}{\end{eqnarray}}
\newcommand{\bi}{\begin{itemize}} \newcommand{\ei}{\end{itemize}}
\newcommand{\bn}{\begin{enumerate}} \newcommand{\en}{\end{enumerate}}
 
\newcommand{\ie}{\emph{i.e., }}
 
\newcommand{\etal}{\emph{et~al.\ }} 
\newcommand{\cf}{cf.\ }
 
\newcommand{\ket}[1]{\ensuremath{|{#1\rangle}}} 
\newcommand{\bra}[1]{\ensuremath{{\langle #1}|}}
\newcommand{\braket}[2]{\ensuremath{\langle #1 | #2 \rangle}}
\newcommand{\ketbra}[2]{\ensuremath{| #1 \rangle \langle #2 |}}

\newcounter{romnum}
\newenvironment{romlist}{\begin{list}{(\roman{romnum})}
{\usecounter{romnum}\setlength{\topsep}{1pt}
\setlength{\itemsep}{1pt}}\rm}{\end{list}}

\newcounter{bracketnum}
\newenvironment{bracketlist}{\begin{list}{(\arabic{bracketnum})}
{\usecounter{bracketnum}\setlength{\topsep}{1pt}
\setlength{\itemsep}{1pt}}\rm}{\end{list}}

\begin{document}

\title{On Zurek's derivation of the Born rule} 

\author{Maximilian Schlosshauer}
\email{MAXL@u.washington.edu}
\affiliation{Department of Physics, University of Washington, Seattle, Washington
    98195}

\author{Arthur Fine}
\email{afine@u.washington.edu} 
\affiliation{Department of Philosophy, University of Washington,
  Seattle, Washington 98195}


\begin{abstract}
  Recently, W.~H.~Zurek presented a novel derivation of the Born rule
  based on a mechanism termed environment-assisted invariance, or
  ``envariance'' (W.~H.~Zurek, \emph{Phys.\ Rev.\ Lett.\ 
  }\textbf{90}(2), 120404 (2003)). We review this approach and
  identify fundamental assumptions that have implicitly entered into
  it, emphasizing issues that any such derivation is likely to face.
\end{abstract}

\maketitle

\section{Introduction}

In standard quantum mechanics, Born's rule \cite{Born:1926:pa} is
simply postulated. A typical formulation of this rule reads:
{\small \begin{quote}
  If an observable $\widehat{O}$, with eigenstates $\{ \ket{o_i} \}$
  and spectrum $\{ o_i \}$, is measured on a system described by the
  state vector $\ket{\psi}$, the probability for the measurement to
  yield the value $o_i$ is given by $p(o_i) = | \braket{o_i}{\psi}
  |^2$.
\end{quote}}
Born's rule is of paramount importance to quantum mechanics as it
introduces a probability concept into the otherwise deterministic
theory and relates it mathematically to the Hilbert space formalism.
No violation of Born's rule has ever been discovered
experimentally---which has certainly supported the r\^ole of the Born
rule as the favorite ingredient of what has been nicknamed the ``shut
up and calculate'' interpretation of quantum mechanics. (Although often
attributed to Feynman, it appears that the nickname was actually
coined by David Mermin \cite{Mermin:1989:tv}. For an example of
such a stance, see \cite{Fuchs:2000:az}).

Replacing the postulate of Born's rule by a derivation would be a
highly desirable goal within quantum theory in general.  The famous
theorem of Gleason \cite{Gleason:1957:zp} presented a mathematical
motivation for the form of the Born probabilities by showing that if
one would like to assign a non-negative real valued function $p(v)$ to
every vector $v$ of a vector space $\mathcal{V}$ of dimension greater
than two such that for every orthonormal basis $\{ v_1, \hdots, v_n
\}$ of $\mathcal{V}$ the sum of the $p(v_i)$ is equal to one,
\begin{equation} \label{eq:gleason}
\sum_i p(v_i) = 1,
\end{equation}
then the only possible choice is $p(v) = | \braket{v}{w} |^2$ for all
vectors $v$ and an arbitrary but fixed vector $w$, provided that
probabilities are assumed to be non-contextual. The normalization
requirement of Eq.~\eqref{eq:gleason} for $p(v)$ with respect to any
orthonormal basis can be physically motivated by remembering that any
orthonormal basis $\{ v_1, \hdots, v_n \}$ can be viewed as the
eigenbasis of observables $\widehat{O} = \sum_i \lambda_i
\ketbra{v_i}{v_i}$, and by referring to the fact that in every
measurement of such an observable $\widehat{O}$ on a system with state
vector $w$ one outcome (represented by the eigenvalue $\lambda_j$
corresponding to one of the eigenvectors $v_j$) will occur, such that
$p(v_j)=1$ and $p(v_i)=0$ for $i \not= j$, and Eq.~\eqref{eq:gleason}
follows. In spite of its mathematical elegance, Gleason's theorem is
usually considered as giving rather little physical insight into the
emergence of quantum probabilities and the Born rule.

Other attempts towards a consistent derivation of the Born
probabilities have previously been made in particular in the context
of relative-state interpretations where both the meaning of
probabilities and their relation to Born's rule requires explicit
elucidation (see, for example,
\cite{Everett:1957:rw,Hartle:1968:gg,DeWitt:1971:pz,%
Graham:1973:ww,Geroch:1984:yt,Farhi:1989:uh}),
but the success of these approaches is controversial
\cite{Stein:1984:uu,Kent:1990:nm,Squires:1990:lz}. A widely disputed
derivation of the Born rule that is solely based on the
non-probabilistic axioms of quantum mechanics and on classical
decision theory (and that is more physically motivated than Gleason's
argument) has been proposed by Deutsch~\cite{Deutsch:1999:tz}. It was
critized by Barnum \etal \cite{Barnum:2000:oz} but was subsequently
defended by Wallace~\cite{Wallace:2003:zr} and put into an operational
framework by Saunders~\cite{Saunders:2002:tz}; no decisive conclusion
seems to have been reached on the success of these derivations thus
far.

A novel and interesting proposal towards a derivation of Born's rule
has recently been put forward by Zurek~\cite{Zurek:2003:rv} (see also
the follow-ups in \cite{Zurek:2003:pl,Zurek:2002:ii}).  Zurek is a key
figure in the development of the decoherence program (for a recent
survey of the program and further references, see
\cite{Zurek:2002:ii,Schlosshauer:2003:ms}) that is based on a study of
open quantum systems and their interaction with the many degrees of
freedom of their environment, leading to explanations for the
emergence of the ``classical'' world of our observation.  However, one
of the remaining loopholes in a consistent derivation of classicality
from decoherence and standard non-collapse quantum mechanics alone has
been tied to the fact that the formalism of decoherence and its
interpretation rely implicitely on Born's rule, but that decoherence
does not yield an independent motivation for the connection between
the quantum mechanical state space formalism and probabilities. Any
derivation of the Born rule from decoherence \cite{Zurek:1998:re} is
therefore subject to the charge of circularity \cite{Zeh:1996:gy}.

To address this criticism, Zurek has suggested a derivation of Born's
rule that is based on the inclusion of the environment---thus matching
well the spirit of the decoherence program---but without relying on
the key elements of decoherence that presume Born's rule and would
thus render the argument circular.  Zurek's derivation is of course
not only relevant in the context of the decoherence program.

Because we consider Zurek's approach promising, we would like to bring
out the assumptions that enter into the derivation but have not been
explicitely mentioned in
Refs.~\cite{Zurek:2003:rv,Zurek:2003:pl,Zurek:2002:ii}.  Hopefully
such an analysis will help in a careful evaluation of the question to
what extent Zurek's derivation can be regarded as fundamental. In
fact, after this paper had been posted online as a preprint, two other
discussions of Zurek's argument have appeared that also describe 
variants of the proof \cite{Barnum:2003:yb,Mohrhoff:2004:tv}.
Moreover, Zurek himself \cite{Zurek:2004:yb} has recently revised his
original derivation in a way that addresses several of the issues
raised in this article and that is more explicit about the assumptions
(some designated now as ``facts'') used in his proof. These correspond
to what we identify in the following discussion.

To anticipate, we find that Zurek's derivation is based on at least
the following assumptions:

\begin{bracketlist}

\item The probability for a particular outcome, \ie for the occurence
  of a specific value of a measured physical quantity, is identified
  with the probability for the eigenstate of the measured observable
  with eigenvalue corresponding to the measured value---an assumption
  that would follow from the {\em eigenvalue--eigenstate link}.
  
\item Probabilities of a system~$\mathcal{S}$ entangled with another
  system~$\mathcal{E}$ are a function of the {\em local} properties of
  $\mathcal{S}$ only, which are exclusively determined by the state
  vector of the {\em composite} system $\mathcal{SE}$.

\item For a composite state in the Schmidt form
  $\ket{\psi_\mathcal{SE}} = \sum_k \lambda_k \ket{s_k} \ket{e_k}$,
  the probability for $\ket{s_k}$ is {\em equal} to the probability
  for $\ket{e_k}$.
  
\item Probabilities associated with a system $\mathcal{S}$ entangled
  with another system $\mathcal{E}$ remain {\em unchanged} when
  certain transformations (namely, Zurek's ``envariant
  transformations'') are applied that only act on $\mathcal{E}$ (and
  similarly for $\mathcal{S}$ and $\mathcal{E}$ interchanged).

\end{bracketlist}

Our paper is organized as follows. First, we review Zurek's derivation
of the Born rule as given in his original papers
\cite{Zurek:2003:rv,Zurek:2003:pl,Zurek:2002:ii}, and also include a
line of reasoning presented in his recent follow-up
\cite{Zurek:2004:yb} (that in turn takes issues raised in the
following discussion into account). We then elucidate and discuss step
by step the assumptions that we believe have entered into Zurek's
approach. In the final section, we summarize our main points.

\section{Review of Zurek's derivation}

\paragraph*{(I)} 
Zurek suggests a derivation of Born's rule for the following pure
state that describes an entanglement between a system $\mathcal{S}$,
described by a Hilbert space $\mathcal{H}_\mathcal{S}$, and its
environment $\mathcal{E}$, represented by a Hilbert space
$\mathcal{H}_\mathcal{E}$:
\begin{equation} \label{eq:state1}
\ket{\psi_\mathcal{SE}} = \sum_k \lambda_k \ket{s_k} \ket{e_k}, 
\end{equation}
where $\{ \ket{s_k} \}$ and $\{ \ket{s_k} \}$ are orthonormal bases of
$\mathcal{H}_\mathcal{S}$ and $\mathcal{H}_\mathcal{E}$, respectively.
Zurek holds that after the $\mathcal{SE}$ correlation has been
established, the system no longer interacts with the environment
\cite[p.~10]{Zurek:2003:pl}, \ie that $\mathcal{E}$ is ``dynamically
decoupled'' \cite[p.~120404-1]{Zurek:2003:rv} and thus ``causally
disconnected'' \cite[p.~754]{Zurek:2002:ii} from $\mathcal{S}$. 

For the sake of clarity and simplicity, we shall in the
following restrict ourselves to the case of coefficients of equal
magnitude and to two-dimensional state spaces
$\mathcal{H}_\mathcal{S}$ and $\mathcal{H}_\mathcal{E}$, \ie we
consider the state
\begin{equation} \label{eq:state2} 
\ket{\psi_\mathcal{SE}} = \frac{1}{\sqrt{2}}
\bigg( e^{i\alpha_1} \ket{s_1} \ket{e_1} + e^{i\alpha_2} \ket{s_2}
\ket{e_2} \bigg).
\end{equation}
Once a valid derivation of Born's rule is accomplished for this
situation, the case of non-equal probabilities and of state spaces of
more than two dimensions can be treated by means of a relatively
straightforward counting argument \cite{Zurek:2003:rv} (at least for
probabilities that are rational numbers). What Zurek's derivation now
aims to establish is the result that for an observer of $\mathcal{S}$,
the probabilities for $\ket{s_1}$ and $\ket{s_2}$ will be equal. That
claim is the focus of our analysis.

\paragraph*{(II)} 
Zurek considers pairs of unitary transformations
$\widehat{U}_\mathcal{S} = \widehat{u}_\mathcal{S} \otimes
\widehat{I}_\mathcal{E}$ and $\widehat{U}_\mathcal{E} =
\widehat{I}_\mathcal{S} \otimes \widehat{u}_\mathcal{E}$. Here
$\widehat{u}_\mathcal{S}$ acts only on the Hilbert state space
$\mathcal{H}_\mathcal{S}$ of $\mathcal{S}$, and
$\widehat{I}_\mathcal{E}$ is the identity operator in
$\mathcal{H}_\mathcal{E}$. Similarly $\widehat{u}_\mathcal{E}$ acts
only on the Hilbert state space $\mathcal{H}_\mathcal{E}$ of
$\mathcal{E}$, and $\widehat{I}_\mathcal{S}$ is the identity operator
in $\mathcal{H}_\mathcal{S}$.

If the composite state $\ket{\psi_\mathcal{SE}}$ is invariant under
the combined application of $\widehat{U}_\mathcal{S}$ and
$\widehat{U}_\mathcal{E}$,
\begin{equation} \label{eq:envar}
\widehat{U}_\mathcal{E} (\widehat{U}_\mathcal{S} 
\ket{\psi_\mathcal{SE}}) = \ket{\psi_\mathcal{SE}},
\end{equation}
the composite state is called \emph{envariant under
  $\widehat{u}_\mathcal{S}$}. (The word ``envariant'' stems from the
abbreviation ``envariance'' of the term ``environment-assisted
invariance'', an expression coined by Zurek). Zurek gives the
following interpretation of envariance
\cite[p.~120404-1]{Zurek:2003:rv}:
{\small \begin{quote} 
    When the transformed property of the system can
    be so ``untransformed'' by acting only on the environment, it is
    not the property of $\mathcal{S}$.  Hence, when $\mathcal{SE}$ is
    in the state $\ket{\psi_\mathcal{SE}}$ with this characteristic,
    it follows that the envariant properties of $\mathcal{S}$ must be
    completely unknown.
\end{quote}}
It is difficult to understand just what the term ``property'' refers
to here, since it is the composite state that is transformed and
untransformed, and so the ``properties'' involved would seem to be
features of the state, not of the system. It seems that envariance
under $\widehat{u}_\mathcal{S}$ is taken to imply that an observer who
``in the spirit of decoherence'' \cite[p.~10]{Zurek:2003:pl} only has
access to $\mathcal{S}$ will not be able to determine features of the
combined state that are affected by $\widehat{u}_\mathcal{S}$ (or,
more properly, by $\widehat{U}_\mathcal{S}$). For such an observer a
local description of $\mathcal{S}$ will be independent of these
features, which may depend on a particular decomposition. While this
general description is far from precise, the uses to which Zurek puts
envariance are clear enough.

\paragraph*{(IIa)} 
The first type of an envariant transformation that Zurek considers is
the pair 
\begin{subequations}
\begin{eqnarray} \label{eq:phasetrsf-S}
\widehat{u}_\mathcal{S}^{(\beta_1,\beta_2)} &=& e^{i\beta_1}\ketbra{s_1}{s_1} +
e^{i\beta_2}\ketbra{s_2}{s_2},  \\  \label{eq:phasetrsf-E}
\widehat{u}_\mathcal{E}^{(\beta_1,\beta_2)} &=& 
e^{-i\beta_1}\ketbra{e_1}{e_1} + e^{-i\beta_2}\ketbra{e_2}{e_2}.  
\end{eqnarray}
\end{subequations}
The effect of the first transformation $\widehat{U}_\mathcal{S} =
\widehat{u}_\mathcal{S}^{(\beta_1,\beta_2)} \otimes
\widehat{I}_\mathcal{E}$ is to change the phases associated with the
terms in the Schmidt state, Eq.~\eqref{eq:state2}, that is,
\begin{equation} \label{eq:state3a} 
\widehat{U}_\mathcal{S}\ket{\psi_\mathcal{SE}} = \frac{1}{\sqrt{2}}
\bigg( e^{i(\alpha_1+\beta_1)} \ket{s_1} \ket{e_1} + 
e^{i(\alpha_2+\beta_2)} \ket{s_2} \ket{e_2} \bigg).
\end{equation}
It is easy to see that if one subsequently acts on this state with
$\widehat{U}_\mathcal{E} = \widehat{I}_\mathcal{S} \otimes
\widehat{u}_\mathcal{E}^{(\beta_1,\beta_2)}$, the original
$\ket{\psi_\mathcal{SE}}$ will be restored. Thus,
$\ket{\psi_\mathcal{SE}}$, and in particular the phases associated
with the states in the Schmidt decomposition of
$\ket{\psi_\mathcal{SE}}$, are envariant under the phase
transformation $\widehat{u}_\mathcal{S}^{(\beta_1,\beta_2)}$ given by
Eq.~\eqref{eq:phasetrsf-S}.

In the spirit of Zurek's interpretation of envariance stated above,
this implies that the phases of the Schmidt coefficients are not a
property of $\mathcal{S}$ alone, so that a local description of
$\mathcal{S}$ cannot depend on the phases $\alpha_1$ and $\alpha_2$ in
the composite state $\ket{\psi_\mathcal{SE}}$ of
Eq.~\eqref{eq:state2}. This leads Zurek to the conclusion that also
the probabilities associated with $\mathcal{S}$ must be independent of
these phases, and that it thus suffices to show that equal likelihoods
arise for the state
\begin{equation} \label{eq:state3} 
\ket{\psi_\mathcal{SE}} = \frac{1}{\sqrt{2}}
\bigg( \ket{s_1} \ket{e_1} + \ket{s_2}
\ket{e_2} \bigg).
\end{equation}
We shall therefore use this state in the rest of the argument.

\paragraph*{(IIb)}
 
Another type of envariant transformations relevant to Zurek's
derivation are so-called ``swaps'',
\begin{subequations}
\bea 
&& \widehat{u}_\mathcal{S}^{(1 \leftrightarrow 2)} =
\ketbra{s_1}{s_2} + \ketbra{s_2}{s_1},
\label{eq:swap-s} \\
&& \widehat{u}_\mathcal{E}^{(1 \leftrightarrow 2)} = \ketbra{e_1}{e_2}
+ \ketbra{e_2}{e_1}.
\label{eq:swap-e} 
\eea
\end{subequations}
Application of $\widehat{U}_\mathcal{S} = \widehat{u}_\mathcal{S}^{(1
\leftrightarrow 2)} \otimes \widehat{I}_\mathcal{E}$, with
$\widehat{u}_\mathcal{S}^{(1 \leftrightarrow 2)}$ from
Eq.~\eqref{eq:swap-s}, to the state $\ket{\psi_\mathcal{SE}}$ in
Eq.~\eqref{eq:state3} yields
\begin{equation} \label{eq:swapped-s}
\widehat{U}_\mathcal{S}\ket{\psi_\mathcal{SE}} = \frac{1}{\sqrt{2}}
\bigg( \ket{s_2} \ket{e_1} + \ket{s_1} \ket{e_2} \bigg), 
\end{equation}
\ie the states of the environment $\mathcal{E}$ correlated with the
states of the system $\mathcal{S}$ have been interchanged. This swap
can obviously be undone by a ``counterswap'' $\widehat{U}_\mathcal{E}
= \widehat{I}_\mathcal{S} \otimes \widehat{u}_\mathcal{E}^{(1
  \leftrightarrow 2)}$, with $\widehat{u}_\mathcal{E}^{(1
  \leftrightarrow 2)}$ from Eq.~\eqref{eq:swap-e}, applied to the
state $\widehat{U}_\mathcal{S}\ket{\psi_\mathcal{SE}}$ in
Eq.~\eqref{eq:swapped-s}. Thus, the composite state
$\ket{\psi_\mathcal{SE}}$, Eq.~\eqref{eq:state3}, is envariant under
swaps. The invariant property is then ``$\ket{s_k}$ is correlated with
$\ket{e_l}$''. On the basis of the interpretation of envariance quoted
above, this implies that a local description of $\mathcal{S}$ must be
independent of which particular environmental state $\ket{e_l}$ is
correlated with a given $\ket{s_k}$, \ie that swapping of the states
of the system cannot be detected by a local observation of
$\mathcal{S}$ alone.

\paragraph*{(IIIa)} 
To make the connection between envariance of $\ket{\psi_\mathcal{SE}}$
under swaps with quantum probabilities and Born's rule, Zurek states
\cite[p.~120404-2]{Zurek:2003:rv}:
{\small \begin{quote} 
    Let us now make a
    rather general (and a bit pedantic) assumption about the measuring
    process: When the states are swapped, the corresponding
    probabilities get relabeled ($i \leftrightarrow j$). This leads us
    to conclude that the probabilities for any two envariantly
    swappable $\ket{s_k}$ are equal.
\end{quote}}
This argument assumes that the swapping transformation (that
interchanges the correlations between the states of the system and the
environment) also swaps the probabilities associated with the states
of the system.  

To motivate this assumption, the following line of reasoning has been
described to us by Zurek in private communication and has subsequently
also appeared in published form in
Refs.~\cite{Barnum:2003:yb,Zurek:2004:yb}. Let $p(\ket{s_1};
\ket{\psi_\mathcal{SE}})$ denote the probability for $\ket{s_1}$ when
the $\mathcal{SE}$ combination is in the composite state
$\ket{\psi_\mathcal{SE}}$, and similarly for $\ket{s_2}$, $\ket{e_1}$
and $\ket{e_2}$.  Before the first swap, Zurek states that
\be \label{eq:1} 
\begin{split}
p(\ket{s_1}; \ket{\psi_\mathcal{SE}}) &= p(\ket{e_1};
\ket{\psi_\mathcal{SE}}), \\ p(\ket{s_2}; \ket{\psi_\mathcal{SE}}) &=
p(\ket{e_2}; \ket{\psi_\mathcal{SE}}), 
\end{split}
\ee
by referring to the direct connection between the states of
$\mathcal{S}$ and $\mathcal{E}$ in the state vector expansion
Eq.~\eqref{eq:state2}. After the first swap (acting on $\mathcal{S}$),
\be \label{eq:2} 
\begin{split}
p(\ket{s_1}; \widehat{U}_\mathcal{S}\ket{\psi_\mathcal{SE}}) &= p(\ket{e_2};
\widehat{U}_\mathcal{S}\ket{\psi_\mathcal{SE}}), \\ p(\ket{s_2};
\widehat{U}_\mathcal{S}\ket{\psi_\mathcal{SE}}) &= p(\ket{e_1};
\widehat{U}_\mathcal{S}\ket{\psi_\mathcal{SE}}), 
\end{split}
\ee
where we have used $\widehat{U}_\mathcal{S}\ket{\psi_\mathcal{SE}}$
instead of $\ket{\psi_\mathcal{SE}}$ as the second argument of the
probability function to take into account the transformation of the
state of $\mathcal{SE}$.  Zurek now holds that under a swap,
properties of the environment cannot have been affected by the first
swap acting on the system $\mathcal{S}$ only, so that we must have
\be \label{eq:3} 
\begin{split}
p(\ket{e_1};
\widehat{U}_\mathcal{S}\ket{\psi_\mathcal{SE}}) &= p(\ket{e_1};
\ket{\psi_\mathcal{SE}}), \\ p(\ket{e_2};
\widehat{U}_\mathcal{S}\ket{\psi_\mathcal{SE}}) &= p(\ket{e_2};
\ket{\psi_\mathcal{SE}}).  
\end{split}
\ee
After the application of the counterswap, we get
\be \label{eq:4} 
\begin{split}
p(\ket{s_1}; \widehat{U}_\mathcal{E}
\widehat{U}_\mathcal{S} \ket{\psi_\mathcal{SE}}) &= p(\ket{e_1};
\widehat{U}_\mathcal{E} \widehat{U}_\mathcal{S}
\ket{\psi_\mathcal{SE}}), \\ p(\ket{s_2}; 
\widehat{U}_\mathcal{E} \widehat{U}_\mathcal{S}
\ket{\psi_\mathcal{SE}}) &= p(\ket{e_2}; \widehat{U}_\mathcal{E} \widehat{U}_\mathcal{S}
\ket{\psi_\mathcal{SE}}), 
\end{split}
\ee
where Zurek has that
\be \label{eq:5} 
\begin{split}
p(\ket{s_1}; \widehat{U}_\mathcal{E}
\widehat{U}_\mathcal{S} \ket{\psi_\mathcal{SE}}) &= p(\ket{s_1};
\widehat{U}_\mathcal{S} \ket{\psi_\mathcal{SE}}), \\ p(\ket{s_2};
\widehat{U}_\mathcal{E} \widehat{U}_\mathcal{S}
\ket{\psi_\mathcal{SE}}) &= p(\ket{s_2}; \widehat{U}_\mathcal{S}
\ket{\psi_\mathcal{SE}}), 
\end{split}
\ee
since the counterswap only acted on $\mathcal{E}$. Moreover, since
after the counterswap the final state vector will be identical to the
initial state vector, Zurek concludes that
\be \label{eq:6}
\begin{split}
  p(\ket{s_1}; \widehat{U}_\mathcal{E} \widehat{U}_\mathcal{S}
  \ket{\psi_\mathcal{SE}}) &= p(\ket{s_1}; \ket{\psi_\mathcal{SE}}), \\
  p(\ket{s_2}; \widehat{U}_\mathcal{E} \widehat{U}_\mathcal{S}
  \ket{\psi_\mathcal{SE}}) &=
  p(\ket{s_2}; \ket{\psi_\mathcal{SE}}),  \\
  p(\ket{e_1}; \widehat{U}_\mathcal{E} \widehat{U}_\mathcal{S}
  \ket{\psi_\mathcal{SE}}) &= p(\ket{e_1}; \ket{\psi_\mathcal{SE}}), \\
  p(\ket{e_2}; \widehat{U}_\mathcal{E} \widehat{U}_\mathcal{S}
  \ket{\psi_\mathcal{SE}}) &= p(\ket{e_2}; \ket{\psi_\mathcal{SE}}).
\end{split}
\ee
This implies, from the above Eqs.~\eqref{eq:1}--\eqref{eq:6}, that
\bea \label{eq:7} 
p(\ket{s_1}; \ket{\psi_\mathcal{SE}}) &=&
p(\ket{s_1}; \widehat{U}_\mathcal{E} \widehat{U}_\mathcal{S}
\ket{\psi_\mathcal{SE}}) = p(\ket{s_1}; \widehat{U}_\mathcal{S}
\ket{\psi_\mathcal{SE}}) \nonumber \\ &=& p(\ket{e_2}; \ket{\psi_\mathcal{SE}})
= p(\ket{s_2}; \ket{\psi_\mathcal{SE}}), 
\eea
which establishes the desired result $p(\ket{s_1};
\ket{\psi_\mathcal{SE}})=p(\ket{s_2}; \ket{\psi_\mathcal{SE}})$.  

\paragraph*{(IIIb)} 
Since the connection between envariance under swaps and equal
probabilities is the crucial step in the derivation, we would like to
mention another line of argument found in Zurek's papers
\cite{Zurek:2003:pl,Zurek:2002:ii} that more explicitely connects
envariance with ignorance and the information available to a local
observer. Here, Zurek considers a von Neumann measurement carried out
on the composite state vector $\ket{\psi_\mathcal{SE}}$ by an
observer, described by ``memory states'' $\ket{\mu_0}$ (the
premeasurement memory state) and $\ket{\mu_1}$, $\ket{\mu_2}$ (the
post-measurement memory states corresponding to the perception of the
``outcomes'' $\ket{s_1}$ and $\ket{s_2}$, respectively):
\bea 
\label{eq:mem} \ket{\mu_0} \ket{\psi_\mathcal{SE}} &\propto&
\ket{\mu_0} \big( \ket{s_1} \ket{e_1} + \ket{s_2} \ket{e_2} \big)\nonumber \\ 
&\longrightarrow& \ket{\mu_1} \ket{s_1} \ket{e_1} + \ket{\mu_2}
\ket{s_2} \ket{e_2}.  
\eea
Zurek then states \cite[p.~755]{Zurek:2002:ii}:
{\small \begin{quote} 
   [Envariance of $\ket{\psi_\mathcal{SE}}$ under
    swaps] allows the observer (who knows the joint state of
    $\mathcal{SE}$ exactly) to conclude that the probabilities of all
    the envariantly swappable outcomes must be the same. The observer
    cannot predict his memory state after the measurement of
    $\mathcal{S}$ because he knows too much: the exact combined state
    of $\mathcal{SE}$.  (\dots) Probabilities refer to the guess the
    observer makes on the basis of his information before the
    measurement about the state of his memory---the future
    outcome---after the measurement. Since the left-hand side of
    Eq.~\eqref{eq:mem} is envariant under swaps of the system states,
    the probabilities of all the states must be equal.
\end{quote}}
In a different paper, Zurek argues \cite[p.~12]{Zurek:2003:pl}:
{\small \begin{quote} When the state of the observer's memory is not
    correlated with the system, and the absolute values of the
    coefficients in the Schmidt decomposition of the entangled state
    describing $\mathcal{SE}$ are all equal, and $\mathcal{E}$ cannot
    be accessed, the resulting state of $\mathcal{S}$ is
    \emph{objectively invariant} under all \emph{local}
    measure-preserving transformations. Thus, with no need for further
    excuses, probabilities of events $\{\ket{s_k} \}$ must be---prior
    to measurement---equal.
\end{quote} }
Obviously, these arguments appeal to a rather different explanation
for the emergence of equal likelihoods from the envariance of
$\ket{\psi_\mathcal{SE}}$ under swaps than the previously quoted
argument. Now, probabilities are introduced from the point of view of
the observer to account for his lack of knowledge of the
individual state of $\mathcal{S}$, since he has perfect knowledge of
the composite state of $\mathcal{SE}$. Then, so goes Zurek's claim,
since the observer cannot detect the swapping of the possible outcome
states $\ket{s_1}$ and $\ket{s_2}$ of $\mathcal{S}$ before the
measurement, he will regard them as ``equivalent'' and therefore
attach equal likelihoods to them.

\section{Discussion}

\paragraph*{(A)}
First of all, Zurek's derivation intrinsically requires the split of
the total Hilbert space $\mathcal{H}$ into at least two subspaces
$\mathcal{H}_\mathcal{S}$ and $\mathcal{H}_\mathcal{E}$ which are
identified with a system $\mathcal{S}$ and its environment
$\mathcal{E}$, where $\mathcal{S}$ is presumed to have interacted with
$\mathcal{E}$ at some point in the past. The environment is then
responsible for the emergence of probabilities within the system,
similar to the spirit of decoherence where the environment is
responsible for the emergence of subjective classicality within the
system \cite{Zurek:2002:ii,Schlosshauer:2003:ms}.

The obvious question is then to what extent the necessity to include
the environment constitutes a restriction of generality.  Apart from
the problem of how to do cosmology, we might take a pragmatic point of
view here by stating that any observation of the events to which we
wish to assign probabilities will always require a measurement-like
context that involves an open system interacting with an external
observer, and that therefore the inability of Zurek's approach to
derive probabilities for a closed, undivided system should not be
considered as a shortcoming of the derivation.

\paragraph*{(B)}
Secondly, we might wonder whether the choice of the entangled pure
Schmidt state, Eq.~\eqref{eq:state1}, implies a lack of generality in the
derivation. Any two-system composite pure state state can be
diagonalized in the Schmidt form above, so the particular form of the
expansion of $\ket{\psi_\mathcal{SE}}$ implies no loss of generality.
Furthermore, if $\rho_\mathcal{SE}$ were non-pure, it could be made pure
simply by enlarging the space $\mathcal{H}_\mathcal{E}$, which cannot
influence probabilities of $\mathcal{S}$ since $\mathcal{E}$ is
assumed to be dynamically decoupled from $\mathcal{S}$ after the
initial interaction that established the entanglement between
$\mathcal{S}$ and $\mathcal{E}$ \cite{Zurek:2003:pl}. We thus conclude
that once the requirement for openness is acknowledged, the
consideration of the state $\ket{\psi_\mathcal{SE}}$,
Eq.~\eqref{eq:state1}, will suffice for a general derivation of the
Born rule.

\paragraph*{(C)}
Before introducing any probability concept into quantum theory, we
need to define what these probabilities are supposed to be assigned
to. Clearly, from the point of view of observations and measurements,
we would like to assign probabilities to the occurence of the specific
values of the observable $O$ that has been measured, \ie to the
``outcomes''. The eigenvalue--eigenstate link of quantum mechanics
postulates that a system has a value for an observable if and only if
the state of the system is an eigenstate characteristic of that value
(or a proper mixture of those eigenstates). If we consider only a
measurement situation, one way of getting this link is first to assume
that the only possible values are outcomes of measurements and that
those are restricted to the eigenvalues $o_i$ of an operator
$\widehat{O}$ that represents the measured observable $O$.  If one
then assumes the collapse or projection postulate, that after the
measurement the state of the system will be in an eigenstate
$\ket{o_i}$ of $\widehat{O}$, it follows that in the non-degenerate
case (\ie when a certain eigenvalue corresponds only to a single
eigenvector of the operator observable) an outcome $o_i$ (the value of
a physical quantity that appears in a measurement) can be directly
related to the eigenstate $\ket{o_i}$ of the measured operator
$\widehat{O}$, as the eigenvalue--eigenstate link requires, and we can
talk equivalently about the probability for a certain outcome, or
eigenvalue or eigenstate.

The basis states $\{ \ket{s_1}, \ket{s_2} \}$ and $\{ \ket{e_1},
\ket{e_2} \}$ appearing in the composite Schmidt state
$\ket{\psi_\mathcal{SE}}$ of Eq.~\eqref{eq:state2} may then be thought
of as the eigenstates of operator observables
$\widehat{O}_\mathcal{S}$ and $\widehat{O}_\mathcal{E}$. In this
sense, Zurek's derivation tries to establish that for the state
$\ket{\psi_\mathcal{SE}}$ of Eq.~\eqref{eq:state2}, the outcomes
represented by the eigenvalues $s_1$ and $s_2$ corresponding to the
eigenstates $\ket{s_1}$ and $\ket{s_2}$ of an operator observable
$\widehat{O}_\mathcal{S} = s_1 \ketbra{s_1}{s_1} + s_2
\ketbra{s_2}{s_2}$ are equally likely.  However, in the context of the
relative-state view that Zurek promotes, he never explicitely talks
about observables and instead directly speaks of determining the
``probabilities of events $\{ \ket{s_k} \}$''
\cite[p.~12]{Zurek:2003:pl}. This identifies the probability for the
occurence of a specific value of a measured physical quantity with the
probability for an eigenstate of the measured observable with an
eigenvalue equal to the measured value. That assumption would be
justified by the eigenvalue--eigenstate link, although it does not
require it.

\paragraph*{(D)}
Zurek furthermore assumes that the probabilities of the outcomes
associated with the \emph{individual} states $\{ \ket{s_1},
\ket{s_2}\}$ of $\mathcal{S}$ and $\{ \ket{e_1}, \ket{e_2}\}$ of
$\mathcal{E}$ are functions of the \emph{composite} state vector
$\ket{\psi_\mathcal{SE}}$ only. (Zurek spells out the assumption that
the derivation will be based on the composite state vector but without
direct reference to probabilities: ``Given the state of the combined
$\mathcal{SE}$ expressed in the Schmidt form (\dots) what sort of
invariant \emph{quantum facts} can be known about $\mathcal{S}$?''
\cite[p.~120404-1]{Zurek:2003:rv}.) This assumption about the
functional dependence of the probabilities is certainly reasonable,
especially since Zurek's aim is clearly to derive Born's rule from
within standard quantum mechanics, where the state vector is assumed
to provide a complete description of the physical system. The
assumption, of course, might well be questioned in a hidden variable
or modal interpretation.

But Zurek's argument requires actually a more detailed assumption than
stated so far.  Obviously, since the $\mathcal{SE}$ composition is in
an entangled pure state, there is no individual state vector of
$\mathcal{S}$ alone. But Zurek \emph{infers} the properties of
$\mathcal{S}$ from the \emph{composite} state vector
$\ket{\psi_\mathcal{SE}}$ by studying its properties under envariant
transformations. The idea is to use envariance to deduce statements
about $\mathcal{S}$ alone. The assumption is now that probabilities
are \emph{local} in the sense that the probabilities that an observer
of $\mathcal{S}$ alone can associate with the ``events''
$\ket{s_k}$---following Zurek's identification of outcomes with
eigenstates, \cf our discussion in (C)---only depend on the
\emph{local} properties of $\mathcal{S}$ (\ie those properties that
cannot be affected by envariant transformations).  An analogous
assumption must also be invoked with respect to $\mathcal{E}$ for
Zurek's argument to go through: probabilities for the states
$\ket{e_k}$ of $\mathcal{E}$ are only dependent on the local
properties of $\mathcal{E}$. This is used to infer the crucial
conclusion that probabilities of $\mathcal{S}$ and $\mathcal{E}$ must
be independent of the envariant properties of $\mathcal{SE}$ (\ie
properties of $\mathcal{SE}$ that do not belong to $\mathcal{S}$ or
$\mathcal{E}$ individually).

This locality of probabilities can be related to the decomposition of
the total Hilbert space into the state space of the system and the
state space of the environment, together with the focus of any
observation on the system alone, on which also the whole definition of
envariance relies.  But without having explicitely connected the
Hilbert state space description with the functional dependence of the
probabilities on the state, affirming that probabilities of
$\mathcal{S}$ and $\mathcal{E}$ can only depend on the local
properties of $\mathcal{S}$ and $\mathcal{E}$, respectively, must be
counted as an important additional assumption.

\paragraph*{(E)}
We would also like to point out that Zurek holds that his argument
does not require a causality or locality assumption, but that
reference to envariance suffices (see, for example,
\cite[p.~120404-2]{Zurek:2003:rv}).  Zurek suggests that one could
alternatively argue for the independence of probabilities from an
envariant property of the entangled $\mathcal{SE}$ combination
directly if causality and the impossibility of faster-than-light
signaling is assumed. He claims that a measurable property associated
with $\mathcal{S}$ cannot depend on an envariant property of the
entangled $\mathcal{SE}$ state, since otherwise one could influence
measurable properties of $\mathcal{S}$ by acting on a ``distant''
environment $\mathcal{E}$, and superluminal communication would be
possible. We do not find this argument compelling, since influencing
measurable properties of a system entangled with a distant partner by
locally acting on the partner does not necessarily require the effect
to be instantaneously transmitted to the system. Even if it were, it
is not clear that this would entail any violation of relativistic
no-signaling requirements (witness the Bohm theory!).

Of course, even if the assumption of causality and the impossibility
of faster-than-light signaling indeed justified the conclusion that
envariant properties of $\mathcal{SE}$ cannot influence locally
measurable physical quantities of $\mathcal{S}$, any necessity for an
appeal to causality to justify Eq.~\eqref{eq:3} would be rather
undesired, since the goal is to derive the Born rule from quantum
theory alone which, strictly speaking, does not entail the
impossibility of superluminal communication. Zurek is clearly aware of
this point by stating that causality is ``more potent''
\cite[p.~754]{Zurek:2002:ii} and ``more `costly' (and not entirely
quantum)'' \cite[p.~120404-2]{Zurek:2003:rv} than envariance. He
consequently holds that his derivation only requires envariance,
although he sometimes seems to implicitely refer to causality, for
instance in arguing that ``only the absolute values of the
coefficients can matter since phases [of the coefficients] can be
altered by acting on $\mathcal{E}$ alone, and $\mathcal{E}$ is
causally disconnected from $\mathcal{S}$''
\cite[p.~10]{Zurek:2003:pl}.

\paragraph*{(F)}
Let us now turn to the chain of relations between the probabilities as
established in Eqs.~\eqref{eq:1}--\eqref{eq:7}. 

\paragraph*{(F1)}
The first relations, Eqs.~\eqref{eq:1}, infer equal probabilities for
the outcomes represented by $\ket{s_1}$ and $\ket{e_1}$ from their
correlation in the direct product $\ket{s_1}\ket{e_1}$ as appearing in
the composite state vector $\ket{\psi_\mathcal{SE}}$ in the Schmidt
decomposition, Eq.~\eqref{eq:state3}. From a point of view that
\emph{presupposes} Born's rule, this assumption is of course trivially
fulfilled, since a simple projection yields
\begin{multline} 
| \underbrace{\bra{e_1}\braket{s_1}{s_1}\ket{e_1}}_{=1} +
\underbrace{\bra{e_2}\braket{s_1}{s_1}\ket{e_1}}_{=0} |^2 \\ = |
\underbrace{\bra{e_1}\braket{s_1}{s_1}\ket{e_1}}_{=1} +
\underbrace{\bra{e_1}\braket{s_2}{s_1}\ket{e_1}}_{=0} |^2, 
\end{multline}
due to orthonormality of the Schmidt basis states $\{ \ket{s_1},
\ket{s_2} \}$ and $\{ \ket{e_1}, \ket{e_2} \}$. But without this
(obviously undesired) presupposition, the relations in
Eqs.~\eqref{eq:1} represent an additional assumption about the
connection between state and probabilities which does not follow from
the assumption (D) that probabilities are a function of the state
vector only.

Of course Eqs.~\eqref{eq:1} may seem innocuous because most of us are
accustomed to thinking in terms of state space projections, and we
make an intuitive connection to probabilities from such projections.
But it seems important in evaluating a derivation of the quantum
probability concept and Born's rule to be aware of where such
presupposed conceptions enter, as assumptions, into the derivation.

\paragraph*{(F2)}
Yet another important assumption appears to be contained in
Eqs.~\eqref{eq:3}. We recall that Zurek justified the relations
$p(\ket{e_1}; \widehat{U}_\mathcal{S}\ket{\psi_\mathcal{SE}}) =
p(\ket{e_1}; \ket{\psi_\mathcal{SE}})$ and $p(\ket{e_2};
\ket{\psi_\mathcal{SE}}) = p(\ket{e_2};
\widehat{U}_\mathcal{S}\ket{\psi_\mathcal{SE}})$ by saying that the
probabilities associated with the environment $\mathcal{E}$ cannot
change as a result of the envariant swap acting on $\mathcal{S}$ since
this swap cannot affect properties of $\mathcal{E}$. An analogous
statement is made in justifying the relations of Eqs.~\eqref{eq:5}.

But this argument requires the assumption that the probabilities
behave similar to the envariant property that the transformation
refers to; \ie, that the behavior of probabilities under envariant
transformations, in particular swaps, is somehow known. This
knowledge, however, is not established by Zurek's derivation, and we
do not see how it could automatically follow from envariance (as
suggested by Zurek). To illustrate this point, consider the following
two statements regarding the implications derived from envariance in
the course of Zurek's argument:

\begin{romlist}
\item Phase envariance implies that the probabilities of $\mathcal{S}$
  must be independent of the phases of the Schmidt coefficients.
\item Envariance under swaps implies that the probabilities of
  $\mathcal{S}$ cannot be influenced by a swap acting on $\mathcal{E}$
  alone.
\end{romlist}

The important difference between (i) and (ii) is that (i) aims at
demonstrating the independence of the probabilities from an envariant
property, whereas (ii) claims invariance of the probabilities under an
envariant transformation. Statement (i) only requires assumption (D)
to hold: phase envariance implies that a local description of
$\mathcal{S}$ cannot depend on the phase factors of the Schmidt
coefficients, so if we assume that the probabilities are a function of
the local properties ascribed to $\mathcal{S}$ on the basis of the
entangled state vector only (and a study of its envariant
transformations), (i) follows. But (ii) requires more: Employing a
reasoning analogous to (i), envariance of the composite state under
swaps solely means that probabilities of $\mathcal{S}$ will not depend
on whether $\ket{s_1}$ is entangled with $\ket{e_1}$ or with
$\ket{e_2}$, since this ``property'' of a specific correlation is not
property of $\mathcal{S}$ alone; but we have said nothing about
whether the application of the swap operation itself to $\mathcal{E}$
might disturb the probabilities associated with $\mathcal{S}$.

We might reinforce this concern by drawing attention to the physical
interpretation of the swap operation. A swap applied to $\mathcal{S}$
implies that the existing correlations $\ket{s_1}\ket{e_1}$ and
$\ket{s_2}\ket{e_2}$ between the system $\mathcal{S}$ and the
environment $\mathcal{E}$ need to be ``undone'', and new correlations
of the form $\ket{s_1}\ket{e_2}$ and $\ket{s_2}\ket{e_1}$ between
$\mathcal{S}$ and $\mathcal{E}$ have to be created. From the form of
the swap transformations, $\widehat{U}_\mathcal{S} =
\widehat{u}_\mathcal{S} \otimes \widehat{I}_\mathcal{E}$ and
$\widehat{U}_\mathcal{E} = \widehat{I}_\mathcal{S} \otimes
\widehat{u}_\mathcal{E}$, it is clear that swaps can be induced by
\emph{local} interactions. But we do not see why shifting features of
$\mathcal{E}$, that is, doing something to the environment, should not
alter the ``guess'' (to use Zurek's expression
\cite[p.~755]{Zurek:2002:ii}; \cf the quote in (IIIb) above) an
observer of $\mathcal{S}$ would make concerning
$\mathcal{S}$-outcomes. Here, if possible, one would like to see some
further argument (or motivation) for why the probabilities of one
system should be immune to swaps among the basis states of the other
system.

\paragraph*{(G)}
Let us finally discuss Zurek's alternative argument based on the
ignorance of an observer of $\mathcal{S}$ with respect to the
individual state of the system $\mathcal{S}$.

In his derivation, Zurek takes the entangled Schmidt state
$\ket{\psi_\mathcal{SE}}$ describing the correlation between
$\mathcal{S}$ and $\mathcal{E}$ as the given starting point and
assumes that the observer somehow knows this state exactly already
\emph{before} any measurement has taken place. According to Zurek this
knowledge seems to imply that the observer is aware of the ``menu'' of
possible outcomes (but cannot attribute a particular outcome state to
$\mathcal{S}$ before the measurement). But since the observer has only
access to $\mathcal{S}$, how is this knowledge established in the
first place? 

In the case of the composite state $\ket{\psi_\mathcal{SE}}$ with
coefficients of equal magnitude, Eq.~\eqref{eq:state2}, one can choose
\emph{any} other orthonormal basis for $\mathcal{H}_\mathcal{S}$ and
always find a corresponding orthonormal basis of
$\mathcal{H}_\mathcal{E}$ such that the composite state
$\ket{\psi_\mathcal{SE}}$ has again the diagonal Schmidt form of
Eq.~\eqref{eq:state2}. Therefore, no preferred basis of
$\mathcal{H}_\mathcal{S}$ or $\mathcal{H}_\mathcal{E}$ has been
singled out. On one hand, this implies that Zurek's argument does not
require any {\em a priori} knowledge of the environmental states
$\ket{e_k}$ for the observer of $\mathcal{S}$. On the other hand,
however, this also means that there is nothing that would tell the
observer of $\mathcal{S}$ which possible ``events'' $\ket{s_k}$ he is
dealing with. (Decoherence provides a mechanism, termed
environment-induced superselection, in which the interaction of
$\mathcal{S}$ with $\mathcal{E}$ singles out a preferred basis in
$\mathcal{H}_\mathcal{S}$ \cite{Zurek:2002:ii}; however, a fundamental
derivation of the Born rule must of course be independent of
decoherence to avoid circularity of the argument.) Even if one holds
that a choice of a particular set of basis vectors is irrelevant to
the derivation since the aim is solely to demonstrate the emergence of
equal likelihoods for \emph{any} orthonormal basis $\{\ket{s_k}\}$,
one is still left with the question how the observer of $\mathcal{S}$
establishes the knowledge that the composite state must be described
by coefficients of equal magnitude.

Zurek then goes on to claim that (i) because all possible outcome
states $\ket{s_k}$ are envariantly swappable, these states appear as
``equivalent'' to the observer of $\mathcal{S}$, and (ii) that this
``equivalence of outcomes'' translates into an attribution of equal
likelihoods for each of these outcomes.  With respect to part (i) of
the argument, perfect knowledge of the pure composite state implies
that the observer (before the measurement) cannot know the individual
state of $\mathcal{S}$, which adds in an ignorance-based probability
concept, but without having established equal likelihoods.  Now, as
mentioned before, envariance under swaps simply means that the
question of which $\ket{e_l}$ of $\mathcal{E}$ is correlated with a
particular $\ket{s_k}$ of $\mathcal{S}$ is irrelevant to a complete
local description of $\mathcal{S}$. But we do not see how this state
of affairs forces the observer of $\mathcal{S}$ to conclude that all
the $\ket{s_k}$ are ``equivalent''. For part (ii), we note that even
if the previous argument did establish an ``equivalence of outcomes'',
this epistemic indifference about the occurrence of a particular
outcome among a set of possible outcomes would not necessarily, from a
general point of view of probability theory, force out the implication
of equal likelihoods; this conclusion would be particularly
questionable when dealing with a set of continuous cardinality.

\section{Concluding remarks}

To summarize, we have pointed out four important assumptions in
Zurek's derivation about the connection between the state vector and
probabilities, and about the behavior of probabilities under envariant
transformations of the state vector:

\begin{bracketlist}

\item The probability for a particular outcome in a measurement is
  directly identified with the probability for an eigenstate of the
  measured observable with an eigenvalue equal to the value of the
  measured physical quantity, an assumption that would follow from
  the eigenvalue--eigenstate link.

\item For two entangled systems $\mathcal{S}$ and $\mathcal{E}$
  described by the Schmidt state $\ket{\psi_\mathcal{SE}} = \sum_k
  \lambda_k \ket{s_k} \ket{e_k}$, probabilities associated with the
  ``outcome states'' $\ket{s_k}$ and $\ket{e_k}$ of each individual
  system are a function of the local properties of the systems only;
  these properties are exclusively determined by the state vector
  $\ket{\psi_\mathcal{SE}}$ of the composite system.
  
\item In an entangled Schmidt state of the form
  $\ket{\psi_\mathcal{SE}} = \sum_k \lambda_k \ket{s_k} \ket{e_k}$,
  the ``outcome states'' $\ket{s_k}$ and $\ket{e_k}$ are equally
  likely: $p(\ket{s_k}; \ket{\psi_\mathcal{SE}}) = p(\ket{e_k};
  \ket{\psi_\mathcal{SE}})$.
    
\item Probabilities associated with the Schmidt states $\ket{s_k}$ of
  a system $\mathcal{S}$ entangled with another system $\mathcal{E}$
  remain unchanged under the application of an envariant
  transformation $\widehat{U}_\mathcal{E} = \widehat{I}_\mathcal{S}
  \otimes \widehat{u}_\mathcal{E}$ that only acts on $\mathcal{E}$
  (and similarly for $\mathcal{S}$ and $\mathcal{E}$ symmetrically
  exchanged): $p(\ket{s_k}; \widehat{U}_\mathcal{E}
  \ket{\psi_\mathcal{SE}}) = p(\ket{s_k}; \ket{\psi_\mathcal{SE}})$
  and $p(\ket{s_k}; \widehat{U}_\mathcal{S} \ket{\psi_\mathcal{SE}}) =
  p(\ket{e_k}; \ket{\psi_\mathcal{SE}})$.

\end{bracketlist}

The necessity for an assumption like (3) in the derivation of the Born
rule can be traced back to a fundamental statement about any
probabilistic theory: We cannot derive probabilities from a theory
that does not already contain some probabilistic concept; at some
stage, we need to ``put probabilities in to get probabilities out''.
Our analysis suggests that this has been done via assumption (3)
above.

We have pointed out that assumption (4) is necessary to have the
argument that is contained in the chain of relations in
Eqs.~\eqref{eq:1}--\eqref{eq:7} between transformed and untransformed
probabilities go through, but we claim that this assumption neither
follows from envariance alone nor from assumption (2). We have also
questioned whether this assumption is physically plausible.

Furthermore we have expressed doubts that Zurek's alternative approach
that appeals to the information available to a local observer is
capable of leading to a derivation of the Born rule. It is neither
clear to us how exact knowledge of the composite state is established
``within'' the local observer before the measurement, nor how
envariance under swaps leads the observer to conclude that the
possible outcomes must be equally likely.

We hope the questions we raise here will not downplay the interest of
Zurek's derivation in the mind of the reader. To the contrary, because
we regard it as significant, we aimed at facilitating a balanced and
careful evaluation of Zurek's approach by bringing out central
assumptions implicit in his derivation. We note that Zurek uses both
``derivation'' and ``motivation'' to describe his treatment of the
emergence of the Born rule.  Once the critical assumptions are made
explicit, however, as they are here and now in his
\cite{Zurek:2004:yb}, the former term seems more appropriate.
Moreover, any derivation of quantum probabilities and Born's rule will
require some set of assumptions that put probabilities into the
theory. In the era of the ``Copenhagen hegemony'', to use Jim
Cushing's apt phrase, probabilities were put in by positing an
``uncontrollable disturbance'' between object and apparatus leading to
a brute quantum ``individuality'' that was taken not to be capable of
further analysis.  Certainly Zurek's approach improves our
understanding of the probabilistic character of quantum theory over
that sort of proposal by at least one quantum leap.

\begin{acknowledgments}

We would like to thank W.~H.~Zurek for thoughtful and helpful
discussions. If we have misinterpreted his approach to the Born rule,
the fault is entirely ours. We are grateful to the referee for useful
comments and suggestions. 

\end{acknowledgments}

\end{document}